\documentclass{nature}

\usepackage{url}
\usepackage{amsmath,amssymb}
\usepackage{graphicx}
\usepackage[varg]{txfonts}
\usepackage{verbatim}
\usepackage{latexsym}
\usepackage{multirow}
\usepackage{fixltx2e}
\usepackage{color}
\usepackage{setspace}
\usepackage{caption}

\newcommand{\mnras}{Mon. Not. Roy. Astron. Soc.}
\newcommand{\apj}{Astrophys. J.}
\newcommand{\aap}{Astron. Astrophys.}
\newcommand{\solphys}{Solar Phys.}

\newcommand{\pnas}{Proc. Nat. Acad. Sci. USA}
\newcommand{\araa}{Annu. Rev. Astron. Astrophys.}

\title{Global-scale equatorial Rossby waves\\ as an essential component of solar internal dynamics} 

\author{Bj\"orn L\"optien$^{1,2}$, Laurent Gizon$^{1,2,3,\ast}$, Aaron~C.~Birch$^1$,Jesper Schou$^1$,Bastian Proxauf$^1$,\\
 Thomas L. Duvall Jr.$^1$, Richard S. Bogart$^4$, Ulrich~R.~Christensen$^1$}

\begin{document}

\maketitle

\begin{affiliations}
\item Max-Planck-Institut f\"ur Sonnensystemforschung, 37077 G\"ottingen, Germany
\item Institut f\"ur Astrophysik, Georg-August-Universit\"at G\"ottingen, 37077 G\"ottingen, Germany
\item Center for Space Science, New York University Abu Dhabi, PO Box 129188, Abu Dhabi, UAE
\item W.W.~Hansen Experimental Physics Laboratory, Stanford University, Stanford, CA 94305, USA
\normalsize{$^\ast$Corresponding author. Email: gizon@mps.mpg.de}
\end{affiliations}

\begin{abstract}
The Sun's complex dynamics is controlled by buoyancy and rotation in the convection zone and by magnetic forces in the atmosphere and corona. While small-scale solar convection is well understood \cite{Stein2012}, the dynamics of large-scale flows in the solar convection zone is not explained by theory or simulations \cite{Hanasoge2012,Gizon2012}. Waves of vorticity due to the Coriolis force, known as Rossby waves \cite{Rossby1939,Rossby1940}, are expected to remove energy out of convection at the largest scales \cite{Vallis1993}. Here we unambiguously detect and characterize retrograde-propagating vorticity waves in the shallow subsurface layers of the Sun at angular wavenumbers below fifteen, with the dispersion relation of textbook sectoral Rossby waves. The waves have lifetimes of several months, well-defined mode frequencies below 200 nHz in a co-rotating frame, and eigenfunctions of vorticity that peak at the equator. Rossby waves have nearly as much vorticity as the convection at the same scales, thus they are an essential component of solar dynamics. We find a transition from turbulence-like to wave-like dynamics around the Rhines scale \cite{Rhines1975} of angular wavenumber of twenty; this might provide an explanation for the puzzling deficit of kinetic energy at the largest spatial scales.
\end{abstract}

Rossby waves, originally observed on Earth as meanders in the polar jet stream \cite{Rossby1939,Rossby1940} and later in ocean currents \cite{Chelton1996}, are the largest-scale planetary waves. They are retrograde-propagating waves of radial vorticity (curl of horizontal flows) with frequencies below twice the Earth’s rotation frequency. Stars are also rotating fluid systems: they may support global-scale Rossby waves, known as r modes in the astrophysical literature \cite{1978MNRAS.182..423P,1981A&A....94..126P,Saio1982,Neiner2012,Saio2017}. The observational search for Rossby waves in stars is motivated by the role that they may play in driving coherent flows such as differential rotation, zonal jets, or long lived vortices \cite{Ward1965,Vallis2006,Liu2011} and in modulating the magnetic field \cite{Gilman1969,Wolff1987}. The Sun, which rotates with a period of approximately 25 days at the equator, is a prime target for Rossby wave investigations as long time series of spatially resolved observations are available. In particular, Ulrich \cite{Ulrich2001} reported the detection of zonal velocity features on the Sun with azimuthal wavenumbers $m \leq 8$ estimated from line-of-sight Doppler observations. These features were measured to persist long enough (seven months) to be solar r-mode candidates. Sturrock et al. \cite{Sturrock2015} reported periodicities in space measurements of the solar radius, which they suggested were due to r modes with $m = 1$. In both these cases however, the flow vorticity was not available for analysis and no dispersion relation was measured.

Generally, the physical nature of a  wave is identified by a well-defined relation between wavenumber and frequency, the dispersion relation.  For a uniformly rotating sphere (angular velocity $\Omega$) and under the approximation of divergence-free and horizontal motion, the dispersion relation of Rossby waves in the rotating frame is $\omega =  - 2m \Omega/[\ell(\ell+1)]$, where $\ell>0$ is the integer spherical harmonic degree and $m$ is the azimuthal order \cite{Saio1982}. The radial vorticity of a mode is described using spherical harmonics and is proportional to $\exp({\rm i} m \phi-{\rm i} \omega t)$ where $\phi$ is longitude increasing in the prograde direction. Without loss of generality we may restrict the values of $m$ to non-negative integers up to $\ell$. Sectoral Rossby waves are obtained by setting $\ell=m$ in this equation; they propagate with a retrograde phase velocity  $\omega/m = -2 \Omega/[m(m+1)] < 0$ and a prograde group velocity $2 \Omega/(m+1)^2 > 0$.  
For the sake of connecting with the geophysics literature, we rephrase these same properties in the beta-plane approximation in the Methods section.

Here we search for global-scale ($m < 20$) vorticity waves in solar fluid flows in the shallow near-surface layers. For $m \sim 10$, solar Rossby waves should have periods longer than four months. Hence, the study of  global-scale Rossby waves on the Sun requires observations covering several years.  

We use a six-year long time-series of intensity images of the solar photosphere from the HMI instrument aboard the SDO spacecraft \cite{2012SoPh..275....3P}. These full-disk images are recorded at a cadence of 45~s and at a spatial resolution that is high enough to resolve the granules, which are prominent convection features of size 1500~km.  Granules extend from the solar surface down to a few 100~km below the solar surface and can be used as tracers of the larger-scale horizontal flows in which they are embedded. 
We derive the two horizontal components of the flow velocity at the solar surface by following the motions of granules between consecutive pairs of HMI intensity images (see Methods). 
The radial vorticity of the flows is then computed and averaged over each solar rotation period in a frame rotating at
\mbox{$\Omega_{\rm ref}/2\pi=453.1$~nHz}, which  is the mean solar  surface equatorial rotation rate for 2010\,--\,2016 inferred from f-mode helioseismology.

Figure~1 shows maps of the horizontal vector velocity (arrows) and the radial component of the vorticity (background colors) for three consecutive solar rotation periods, after applying a Gaussian spatial filter with a standard deviation of $7^\circ$.  These maps exhibit a complex flow pattern with many examples of vortical flow features. Some vorticity features can be followed from one rotation to the next, although the correlation coefficient between successive maps is only $\sim 0.1$. A more narrow low-pass filter with $\ell \leq 12$ results in a larger correlation coefficient of $\sim 0.25$. When only the latitudes between $\pm20^\circ$ are considered, the correlation coefficient is $\sim0.3$.

In order to study the evolution of vorticity features, we select locations of local maxima of vorticity within $20^\circ$ of the equator in a selected reference map (see black squares in {Fig.~1}), and then compute the evolution of the average vorticity feature. This straightforward method has been useful for studying the evolution of intermediate-scale convection  \cite{2015A&A...581A..67L}.  In this process we align the individual vorticity maxima by shifting them in longitude and latitude. Keeping the same locations, we perform the same averaging over the vorticity values in the previous and subsequent  ten rotation periods, to obtain the time evolution of the average features identified in the reference map. To enhance the signal-to-noise ratio, the above averaging procedure is repeated for all possible choices of reference maps, to finally obtain the evolution of the averaged vorticity features shown in {Fig.~2}.  In the figure, we observe a long-lived (up to about seven rotation periods) vorticity pattern consisting of alternating vorticity maxima and minima consistent with a dispersive wave packet. The phase of the pattern propagates in the retrograde direction (up and to the left in the figure), while the envelope of the pattern propagates in the prograde direction (up and to the right). The phase and group velocities are near the expected values of Rossby waves with $m\simeq 9$. At short time lags of a few rotation periods we expect the vorticity signal to include a contribution from convection.

We have also studied the horizontal divergence averaged over the locations of the vorticity features (i.e., locations picked in the same way as the locations shown in Fig. 1). We find that there is no noticeable horizontal divergence associated with the Rossby waves, as expected from simple models (see power spectrum of the divergence in Supplementary Fig.~S1c and S1d).

To further characterize the wavepacket seen in {Fig.~2}, we compute the power spectrum of the radial vorticity from granulation-tracking maps obtained with the full cadence of 30~min. {Figure~3a} shows the $\ell=m$ power spectrum based on a spherical harmonic decomposition of the vorticity maps.  In the region $4 \leq m \leq 15$, we see a narrow ridge of power of width less than 50~nHz ({Fig.~3b}).  The ridge of power  closely follows the classical dispersion relation (black curve) for sectoral Rossby waves, $\omega = -2\Omega_{\rm ref}/(m+1)$. For the sake of intuition, the observed phase velocity of the Rossby waves is approximately $-40 (m/10)^{-2}$~m\,s$^{-1}$ 
and the group velocity has the opposite value. We have not seen evidence for Rossby waves in the power spectra with $m \neq \ell$ (see Supplementary Fig. S1b). Notice that, by construction, the vorticity maps analyzed here do not have power for $|m| \leq 1$ due to field effect corrections; thus we cannot comment on the existence of the $m=1$ modes proposed in ref. \cite{Sturrock2015}.

The Rossby wave frequencies and lifetimes can be measured by fitting Lorentzian profiles to the peaks of power (see {Fig.~3c} for $m=8$ and Fig.~4 for the  other $m$ values between 4 and 15). To extract the parameters of the Rossby waves in frequency space from the power spectra,we fit the peaks with Lorentzian functions \cite{1990ApJ...364..699A,1994A&A...289..649T}. A maximum likelihood technique is used under the assumption that the power at each frequency is distributed according to an exponential distribution. Error estimates are derived using Monte-Carlo simulations: after fitting the data, we generate one thousand realizations of the power spectrum using the parameters of the fit. Fitting these random realizations allows us to estimate the noise in the fit parameters. The linewidth is a free parameter for each mode. The peak frequencies and linewidths resulting from the fits are listed in Table~1. Mode frequencies are plotted in Fig.~3d. We find that most of the measured frequencies are within $\sim 10$~nHz of the classical Rossby wave frequencies, i.e.\ within two standard deviations, with the most significant deviations seen for the modes $m=4$, $m=8$, and $m=13$. Part of this difference may be due to the effect of differential rotation \cite{1998ApJ...502..961W}.  The average $e$-folding lifetime  for the Rossby waves is measured to be about four months (using the linewidths given in Table~1).

Wave amplitude is another important quantity. Since the eigenfunctions of the Rossby waves are not pure spherical harmonics, we estimate wave amplitudes by computing the rms vorticity of  individual modes in a strip in latitude between $\pm 20^\circ$. 
We perform a Fourier transform in longitude and time at each latitude. For each $m$ we select a band in frequency (60~nHz width) centered around the mode frequencies listed in {Table~1} and perform an inverse Fourier transform to revert to longitude and time. This results in a time-series of Rossby-wave vorticity maps for each azimuthal order. From these filtered maps, we then compute the rms vorticity over latitude and longitude in the latitude range between $\pm 20^\circ$ (see {Table~1}). We compute the rms vorticity for each synoptic map and then average the rms vorticity over time. The rms vorticity of the Rossby waves (after summing the filtered maps over $m$, with $m$ between 4 and 15) is $(6.7 \pm 0.2) \times 10^{-8}$~s$^{-1}$. For comparison, the rms vorticity of the convection only is $(1.01 \pm 0.03) \times 10^{-7}$~s$^{-1}$. This indicates that Rossby waves are a very significant component of solar dynamics at these large spatial scales.

Next we estimate the latitudinal variations  of the mode eigenfunctions at each $m$ by computing the covariance between the radial vorticity at the equator and the radial vorticity at all other latitudes (see Fig.~3e). We find that the covariance function peaks at the equator and is confined to low latitudes. In addition, the sign of the covariance is negative at higher  latitudes. For example for $8\le m\le 11$, the sign reversal occurs at latitudes of about $\pm 25^\circ$. As $m$ increases, the latitude of the sign reversal decreases. Because the covariance functions are more confined to low latitudes than the $\ell=m$ spherical harmonics, this suggests that the modes are trapped in the equatorial region.  We note that latitudinal differential rotation may have a significant effect on the horizontal  eigenfunctions \cite{Zhang1989}.

In order to gain some insight into the depth dependence of the Rossby waves, we have constructed maps of radial vorticity at different depths in the solar interior using a technique of helioseismology known as ring-diagram analysis \cite{Bogart2015}, see Methods. {A comparison between the power spectra at depth $0.7$~Mm (helioseismology) and at the surface (granulation tracking) is provided in Fig.~4. Rossby waves are easily identified  in both cases as peaks of excess power at nearly the same frequencies. The scales in wave power are however not directly comparable since helioseismology has lower horizontal  resolution than the granulation-tracking method: wave power drops faster in the helioseismic data as $m$ increases. 
By comparing the vorticity power spectra at three different depths in the solar interior using helioseismology only (Supplementary Figs.~S2 and~S3), we find that the amplitudes of the Rossby waves do not depend strongly on depth throughout the entire depth range that we have analyzed so far (down to $21$~Mm).} A further characterization of the radial eigenfunctions will require helioseismic inferences at greater depths. This, in combination with modeling, will help determine the radial order(s) of the eigenfunctions (number of nodes in the radial direction) \cite{1981A&A....94..126P,Saio1982,1986SoPh..105....1W}. It is possible that several closely-spaced radial orders contribute to the peaks of power that we have observed.

One may ask if there is a connection between the internal Rossby waves and the large-scale traveling patterns of magnetic bright points in the Sun's corona reported by McIntosh et al. \cite{McIntosh2017}. These patterns are seen at active latitudes with an apparent phase velocity of about $3$~m\,s$^{-1}$ (prograde) and an apparent group velocity of about $-24$~m\,s$^{-1}$ (retrograde) measured in the Carrington frame. These observations were interpreted as evidence of ``Rossby-like waves'' or ``magnetic Rossby waves'', although the directions of the phase and group speeds are \textit{opposite} from what is expected for classical Rossby waves. [Note: By language convention, west on the Sun is the prograde direction, while west on the Earth is the retrograde direction.] We do not see any evidence for a connection  between these moving coronal features and the internal Rossby waves described in the present paper.

There are interesting questions about mode physics. Why are only the latitudinally-sym\-me\-tric $m=\ell$ modes clearly visible in our observations (see Supplementary Fig.~1b)? Linear theory under the assumption of purely horizontal (toroidal) flow would also allow for other modes to exist, including latitudinally antisymmetric modes. 
Eigenvalue calculations for neutrally-stratified stellar models by  Yoshida \& Lee \cite{Yoshida2000} found only r modes with  $m=\ell$  and no node in the radial direction.
To add to this theoretical prediction, we carried out simple simulations in a rotating spherical shell of finite depth that show that toroidal flow patterns with $m \neq \ell$ interact with poloidal (radial) flows.  As a result they cascade to increasingly higher $\ell$ values where energy can be dissipated (see Methods and Supplementary Fig.~S4). Toroidal flow patterns with $m=\ell$ decay only at the viscous time scale. The lifetime of sectoral Rossby waves may thus tell us about the turbulent viscosity in the solar interior \cite{1986SoPh..105....1W}, a quantity of interest for convection theory and dynamo models.

Other questions come to mind. What is the role of equatorial Rossby waves in the dynamics of solar large-scale flows?  Non-interacting sectoral Rossby waves cannot transport angular momentum across latitudes and thus cannot affect latitudinal differential rotation. 
Rossby waves presumably interact with convective flows at comparable scales.
Rhines \cite{Rhines1975} suggested that turbulence transitions to larger-scale Rossby waves at the angular  wavenumber where the time scales of the turbulence and the waves match, i.e. at  $\ell_{\rm R} \sim  (\Omega R / v_{\rm turb})^{1/2}  =  20$ where $R$ is the solar radius and the rms turbulence velocity  $v_{\rm turb}$ is chosen to be the observed value of $4$~m\,s$^{-1}$ from Fig.~1. 
The sectoral  Rossby waves discovered here are clearly above the convective background  in the vorticity power spectrum  below  $m\sim 15$ (Fig.~3), above which the waves are difficult to distinguish from convection. This is consistent with the suggestion by Vallis and Maltrud \cite{Vallis1993} , in the context of geostrophic turbulence, that Rossby waves determine the vorticity spectrum at the largest scales.  The solar Rossby waves have nearly as much radial vorticity as the convection for $m \leq 15$.  Accounting for Rossby waves may help resolve the discrepancy between observations and models of the amplitudes of flows in the solar convection zone at large spatial scales \cite{Hanasoge2012, Gizon2012}.

Solar Rossby waves may have important applications beyond solar physics. They are a particular case of inertial waves that have been proposed as an efficient mechanism to transfer angular momentum to a star from an orbiting close-in planet in tidal interaction \cite{Ogilvie2014}. The dissipation of the waves into heat may take energy away from the system and lead to a secular evolution of the orbital parameters. The inefficiency of the tidal dissipation is specified by the ratio, $Q_{\rm tide}$, of the energy in the tide and the energy dissipated per period. The value of $Q_{\rm tide}$ is smallest when associated with a resonant inertial mode: at resonance, it is inversely proportional to the mode quality factor $Q$ \cite{Wu2005b}. The Rossby mode quality factors that we measure for the Sun
(Table~1) thus provide precious observational information to calibrate theories of the dissipation
of tides in Sun-like stars through inertial modes.

\noindent
{\bf Data availability:} 
{The data that support the plots within this paper and other findings of this study are available from the corresponding author upon reasonable request. The HMI data can be downloaded from the Joint Science Operations Center at http://jsoc.stanford.edu/.
The p-mode ring fits are available in the data series hmi.rdVfitsf and the inversion code is expected to be released as version~0.92 (available on request).
}

\clearpage

{\noindent \large \bf References}
\providecommand{\noopsort}[1]{}\providecommand{\singleletter}[1]{#1}%
  \newcommand{\noop}[1]{}

\clearpage

\noindent{\bf Acknowledgments}\\
We thank Robert\,H.\ Cameron, Cilia Damiani,  Hideyuki Hotta, St\'ephane Mathis, Olivier Pauluis, and Andreas Tilgner for useful discussions. The HMI data are courtesy of NASA/SDO and the HMI Science Team. 
The data were processed at the German Data Center for SDO funded by the German Aerospace Center (DLR). L.\,G.\ acknowledges partial research funding from the NYUAD Institute under grant G1502. 
B.\,P.\ is a member of the International Max Planck Research School for Solar System Science at the University of G\"ottingen.\\

\noindent
{\bf Author Contributions} \\
B.\,L., L.\,G., and A.\,C.\,B.\ designed the research. All authors performed research.  B.\,P.\ contributed to the computation of the vorticity maps using ring-diagram analysis. B.\,L., L.\,G., and A.\,C.\,B.\ drafted the paper and all authors contributed to the final manuscript.\\

\noindent
{\bf Competing interests} \\
The authors declare no competing financial interests.\\

\noindent
{\bf Corresponding author}\\
Correspondence to Laurent Gizon (gizon@mps.mpg.de).

\clearpage


\begin{figure}
\includegraphics[width=0.9\textwidth]{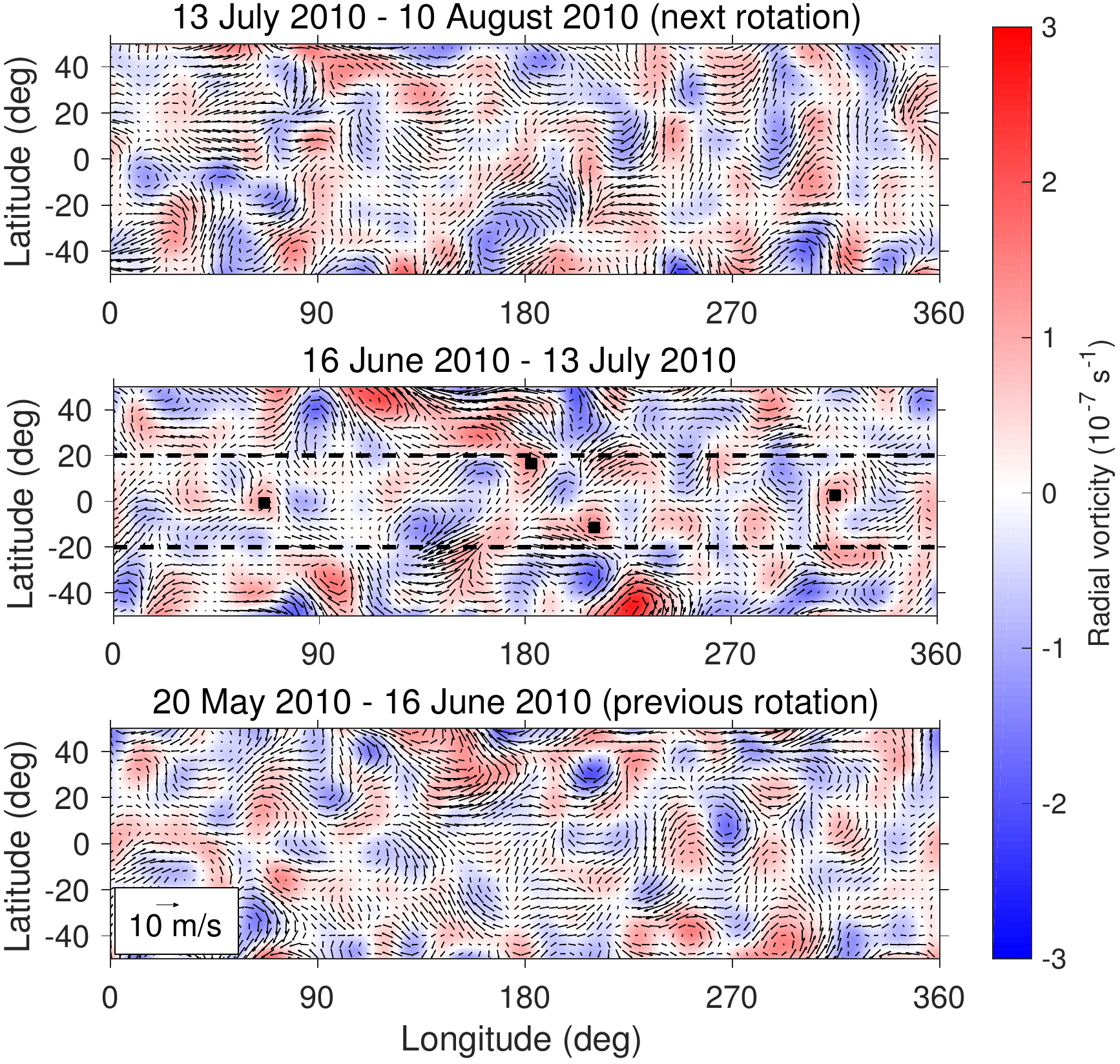}
\caption{
{\bf Surface radial vorticity for three consecutive solar rotations.} 
The radial vorticity is derived using the granulation tracking method (color scale, positive values are anti-clockwise).  The arrows visualize the horizontal flow velocities (see arrow in lower left corner for the scale) measured in a frame rotating at the equatorial surface angular velocity $\Omega_{\rm ref}$. The maps have been smoothed with a Gaussian function (standard deviation of $7^\circ$) in order to remove the contribution from small spatial scales with $\ell > 20$, and the rms velocity of the flows is about 4~m\,s$^{-1}$. The black squares in the middle panel indicate local maxima of the radial vorticity above $10^{-7}$~s$^{-1}$ in the latitude range between $-20^\circ$ and $20^\circ$ (horizontal dashed lines).
}
\label{fig:flow_maps}
\end{figure}

\begin{figure}
\centering
\includegraphics[width=11.5cm]{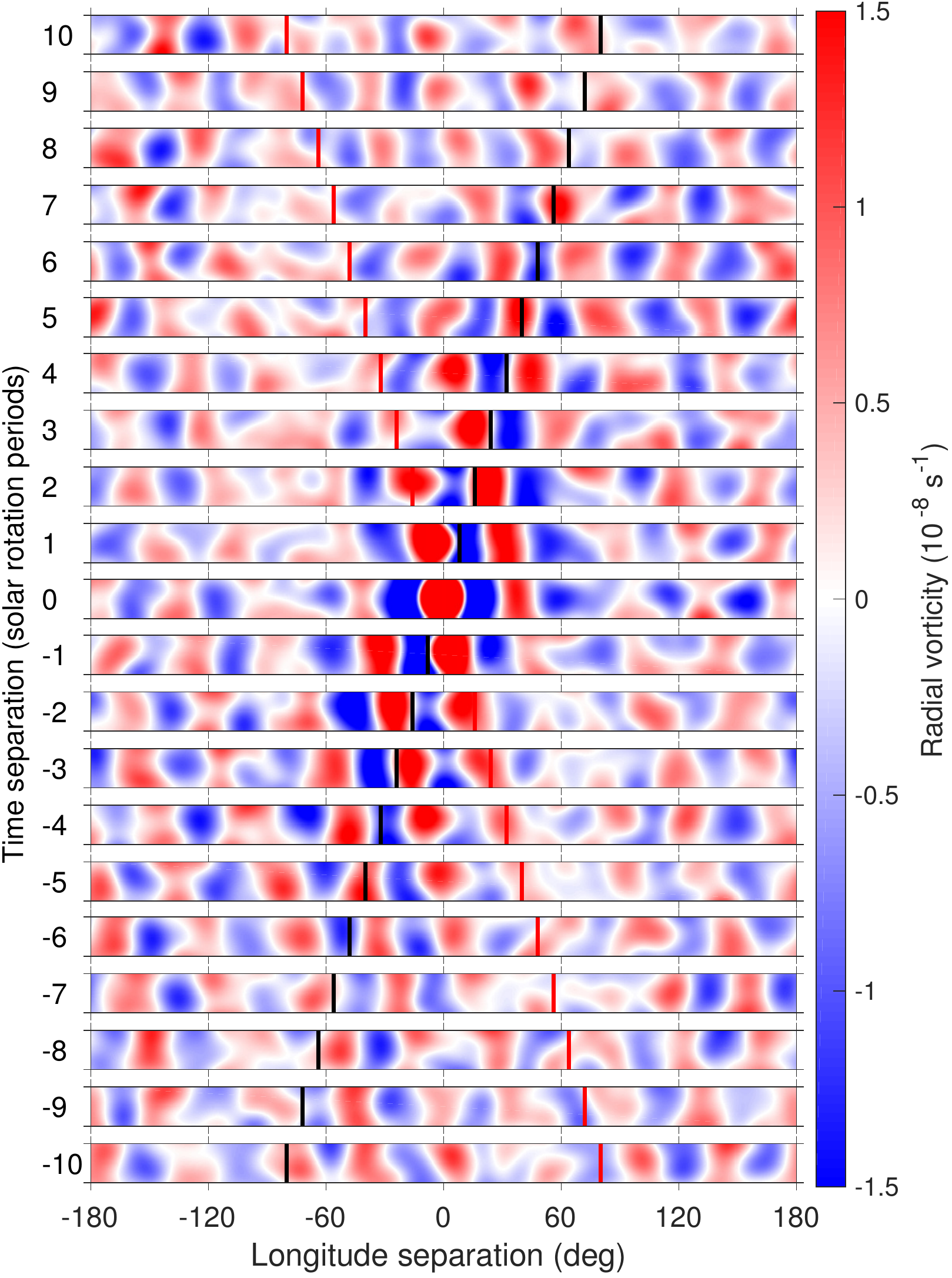}
\caption{
{\bf Longitude-time evolution of the average radial vorticity feature.}
The 305 individual vorticity features were identified by their local maxima, as shown in the middle panel of Fig.~1. The evolution of the average vorticity feature is followed in time over $\pm10$ solar rotation periods, in bands  that cover $360^\circ$ in longitude and $\pm 10^\circ$ in latitude. 
The vorticity values of the average feature range between $-3.8\times10^{-8}$~s$^{-1}$ and  $1.3\times10^{-7}$~s$^{-1}$. In the plot, values are truncated to $\pm 1.5\times 10^{-8}$~s$^{-1}$. 
The red lines indicate the phase speed and the black lines the group speed for an equatorial Rossby wave with $m=9$. As in Fig.~1, the high spatial frequencies were filtered out by convolving the velocity maps by a Gaussian with standard deviation of $7^\circ$.
}
\end{figure}

\begin{figure}
\centering
\includegraphics[width=0.8\textwidth]{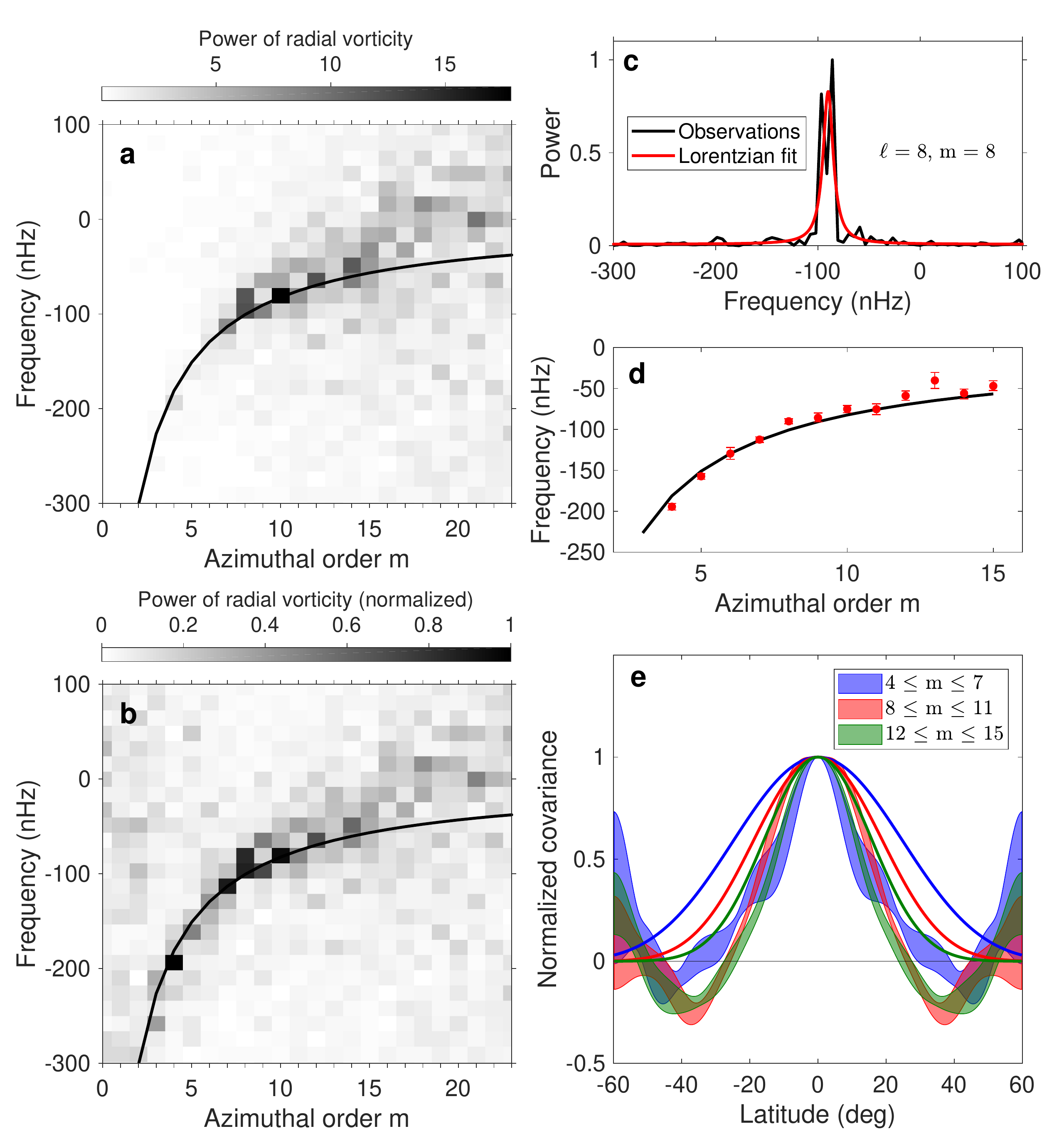}
\caption{{\bf Dispersion relation and horizontal eigenfunctions of equatorial  Rossby waves.}
({\bf a}) Power spectrum of surface radial vorticity from granulation tracking for $m=\ell$  using a spherical harmonic decomposition. The frequency spacing is 16~nHz. The vorticity is measured in a frame rotating at the surface equatorial angular velocity $\Omega_{\rm ref}$. The black curve shows the theoretical dispersion relation for sectoral Rossby waves in the rotating frame. 
({\bf b}) Same as panel (a) except that the power at each $m$ is normalized by its mean over frequencies between $-310$ and $110$~nHz.
({\bf c}) Power spectrum of surface radial vorticity at $\ell=m=8$ (black curve, without smoothing or binning). The red curve shows a Lorentzian fit to the data, used to measure the peak frequency and the linewidth. 
({\bf d}) Measured peak frequencies in the surface power spectrum (red points with one-sigma errors) and theoretical mode frequencies  of classical Rossby waves (black curve). 
({\bf e}) Estimate of the latitudinal eigenfunctions of the Rossby waves derived from the covariance of the surface vorticity at the equator and other latitudes. Averages over three sets of azimuthal orders (see inset) are plotted as shaded areas ($\pm 1 \sigma$ around the mean). The thin lines show averages of the corresponding Legendre polynomials. 
}
\end{figure}

\begin{figure}
\includegraphics[width=0.9\textwidth]{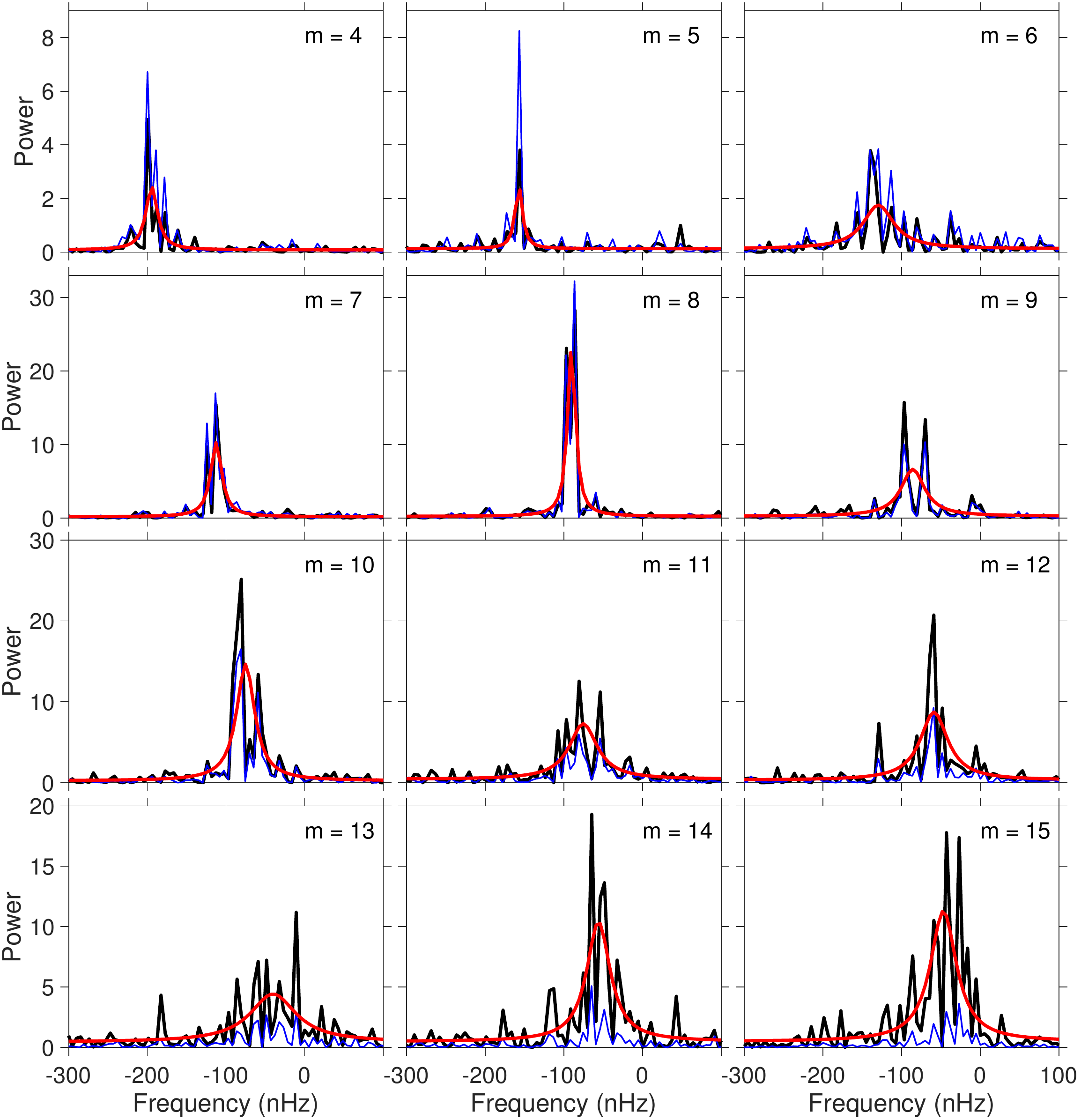}
\caption{{\bf Sectoral power spectra of the radial vorticity for $m$ in the range from 4 to 15.}
The surface power spectra (granulation tracking) are plotted in black. The Lorentzian fits (from which the parameters of Table~1 are extracted) are plotted in red. Note that we cannot exclude
the possibility that several closely-packed radial orders may contribute to the observed peaks of power. The blue thin curves show the power spectra from ring-diagram helioseismology at a depth of $0.7$~Mm. The normalization of the ring-diagram data is such that the power in the $m = 8$ mode at depth $0.7$~Mm is the same as for the surface data.}
\end{figure}

\begin{table}
\centering
\begin{tabular}{cccccc}
  $\boldsymbol{(\ell,m)}$ & $\boldsymbol{\frac{-2\Omega_{\rm ref}}{2\pi(m+1)}}$& {\bf Frequency} & {\bf Linewidth}  & {\bf Q factor} & {\bf Rms vorticity} \\
  {} & {\bf (nHz)} & {\bf (nHz)} &  {\bf (nHz)} &  
  & $\boldsymbol{(10^{-8}~{\rm s}^{-1})}$ \\
  \hline \\
$(4,4)$ & $-181$  & $-194^{+5}_{-4}$ & $18^{+14}_{-7}$ & $10.8$ & $0.75\pm 0.03$\\
$(5,5)$ & $-151$ & $-157\pm 4$ & $11^{+14}_{-6}$ & $14.3$ & $0.94\pm 0.03$\\
$(6,6)$ & $-129$ & $-129\pm 8$ & $47^{+28}_{-16}$ & $2.7$ & $1.23\pm 0.04$\\
$(7,7)$ & $-113$ & $-112\pm 4$ & $17^{+10}_{-7}$ & $6.6$ & $1.57\pm 0.07$\\
$(8,8)$ & $-101$ & $-90\pm 3$ & $12^{+7}_{-5}$ & $7.5$ & $2.16\pm 0.12$ \\
$(9,9)$  & $-91$ & $-86\pm 6$ & $37^{+21}_{-11}$ & $2.3$ & $1.63\pm 0.08$\\
$(10,10)$ & $-82$ & $-75\pm 5$ & $28^{+12}_{-10}$ & $2.7$ & $2.27\pm 0.10$\\
$(11,11)$ & $-76$ & $-75\pm 7$ & $43^{+23}_{-13}$ & $1.7$ & $2.00\pm 0.12$\\
$(12,12)$ & $-70$ & $-59\pm 6$ & $42^{+20}_{-12}$ & $1.4$ & $1.94\pm 0.10$\\
$(13,13)$ & $-65$ & $-40 \pm 10$ & $71^{+38}_{-22}$ & $0.6$ & $1.94\pm 0.09$\\
$(14,14)$ & $-60$ & $-56^{+6}_{-7}$ & $36^{+20}_{-13}$ & $1.6$ & $2.29\pm 0.07$\\
$(15,15)$ & $-57$ & $-47^{+7}_{-6}$ & $40^{+21}_{-12}$ & $1.2$ & $2.42\pm 0.09$
\end{tabular}
\caption{{\bf Parameters of solar sectoral equatorial Rossby waves in the frame rotating at the surface equatorial angular velocity $\boldsymbol{\Omega_{\rm ref}}$.} 
The mode parameters are estimated from Lorentzian fits to the vorticity power spectra for the granulation-tracking data (see Fig.~4). The second column in the table gives the theoretical mode frequencies of Rossby waves in a uniformly rotating solar model. The observed frequencies and linewidths (full widths at half maxima) are inferred using Lorentzian fits to the peaks in the power spectrum. The wave quality factor $Q$ is defined as the ratio between the observed wave frequencies and the linewidths. The vorticity value in the last column is the root mean square of the radial vorticity near the dispersion relation. Error bounds give the 68\% confidence interval.}
\label{tab:freq}
\end{table}

\clearpage

\noindent{\Large \bf Methods}\\

\noindent{\bf Maps of horizontal flows on the solar surface (granulation tracking)}

\noindent We derive the two horizontal components of the flow velocity at the solar surface by following granules in intensity images; granules are advected by larger-scale flows. We use the local correlation tracking method that consists of measuring the spatial shifts from the cross-correlation function between small patches ($\sim4$~Mm diameter) extracted from consecutive intensity images \cite{2004ApJ...610.1148W,2008ASPC..383..373F}. The 45~s cadence of HMI intensity images is much shorter than the evolution time of granules (10~min). The spatial shifts are averaged over pairs of images separated by 30~min and spanning a total of six~years (20~May~2010 to 12~April~2016). The resulting flow maps contain a systematic field effect inherent to the local correlation tracking method known as the ``shrinking Sun effect'' \cite{2016A&A...590A.130L}. We remove this center-to-limb and large-scale effect by projecting the flow maps onto a basis of Zernike polynomials with degrees less than eight and removing the contributions at zero frequency and frequencies related to the orbit of SDO and the orbit of the Earth \cite{Loeptien2017}.  After removal of this effect, we remap the flow maps into a heliographic longitude-latitude coordinate system  in a frame rotating at the equatorial surface rotation rate $\Omega_{\rm ref}$. 
The data from a full rotation period are used  to obtain a synoptic flow map.   At each point on the map, the data are effectively averaged over about eight~days. The radial component of vorticity is obtained by second-order finite differences. This procedure is repeated to analyze six years of observations, from 20~May~2010 to 12~April~2016.\\

\noindent {\bf Maps of horizontal flows below the solar surface (helioseismology)}

\noindent To probe the subsurface layers,
we use flow maps obtained from ring-diagram analysis as provided by the HMI pipeline \cite{Bogart2015}, with only small modifications to the inversion code. The HMI Doppler velocity observations are remapped onto patches that are tracked at the Carrington rate. The patches are tracked for $28.8$~hr, have a size of $15^\circ$, and overlap by 50\% in each direction. The frequencies of the modes of oscillations are extracted from the local power spectra for each patch and are inverted using a one-dimensional optimally-localized-averaging technique to estimate the horizontal flows as a function of depth in the solar interior. {These maps sample the horizontal flows up to a maximum depth of about 21~Mm.}  For comparison with the surface flow maps obtained from granulation tracking, we remap the ring-analysis flow maps into the frame that rotates at the equatorial surface rotation rate $\Omega_{\rm ref}$. We compute the radial vorticity at each depth using second-order finite differences.\\

\noindent {\bf Rossby waves in the beta-plane approximation}

\noindent To connect with the geophysics literature, we recall here the basic properties of Rossby waves in a rotating Cartesian  coordinate system. In the beta-plane approximation \cite{Rossby1939}, Rossby waves of the form $\cos{(k_x x + k_y y - \omega t)}$ have the dispersion relation $\omega = U_0 k_x - \beta k_x / k^2$, where $k_x$ and $k_y$ are the wavenumbers in the prograde and northern directions, $k = (k_x^2 + k_y^2)^{1/2}$ is the horizontal wavenumber,  and $U_0$ is a background zonal flow.   The parameter $\beta$ is a measure of the variation of the Coriolis force with latitude $\lambda$; it is given by $\beta=2\Omega\cos\lambda/R$, where $\Omega$ is the angular velocity of the spherical body with radius $R$. In the frame of the flow $U_0$, Rossby waves with $k_y=0$ are characterized by a retrograde  phase speed $v_p = \omega/k_x = -\beta/k_x^2$ and a prograde  group speed $v_g = \partial\omega/\partial k_x = \beta/k_x^2$. The above dispersion relation is obtained in the shallow-water approximation for an incompressible fluid, which describes a motion that is horizontal and without horizontal divergence \cite{Pedlosky,Rieutord}.\\

\noindent{\bf Incompressible Rossby modes in a spherical shell}

\noindent To understand the special role of Rossby modes with $m=\ell$, we consider an incompressible flow in a rotating spherical shell of finite
thickness $d$ with impenetrable free-slip boundaries.  When neglecting  nonlinear momentum advection  and viscous forces, the flow is governed by
$\partial {\bf u}/\partial t = -2{\bf\Omega}\times{\bf u} - \nabla p/\rho$,
where ${\bf u}$ is velocity, ${\bf \Omega}$ is the rotation vector, $p$ is pressure, and $\rho$ is density.
The (divergence-free) flow field can be represented in terms of a complex-valued toroidal scalar
potential $Z$ and a poloidal potential $W$ that are expanded in series of
spherical harmonic functions,
e.g. $ Z({\bf r}, t) = \sum{\,Z_{\ell m}(r,t) Y_{\ell m}(\theta,\phi)}$.
The velocity is obtained as
${\bf u} = \nabla\times (Z{\bf r}) \, + \, \nabla\times\nabla\times (W{\bf r})$.
In the limit of vanishing shell thickness, i.e. in the absence of poloidal flow,
Rossby modes drifting in longitude are eigensolutions of the above equation and
described by $Z_{\ell m}(r,t) = Z_{\ell m}(r, t=0) \exp{ (-i\omega_{\ell m} t) }$ with
$\omega_{\ell m} = -2\Omega m/[\ell(\ell+1)]$ for arbitrary $\ell$ and $m$.
In a layer of finite depth, the Coriolis force term in the equation couples
a toroidal flow mode $Z_{\ell  m}$ to the poloidal flow modes at the same
order $m$ and the two neighbouring degrees,  $W_{\ell+1,m}$ and $W_{\ell-1,m}$
\cite{Tilgner99}. In turn, a poloidal mode $W_{\ell m}$ is coupled to
$Z_{\ell+1,m}$ and $Z_{\ell-1,m}$. The case of modes $Z_{m m}$ is special
because they couple only to a single neighbouring mode, $W_{m+1,m}$, since
$\ell \ge m$ and modes with $(m-1,m)$ do not exist.
It can be shown that for the correct radial eigenfunction,
$Z_{m m} \propto r^m$, the coupling to $W_{m+1,m}$ becomes zero.
Therefore toroidal modes with $\ell$ = $m$ are eigenmodes of the governing
equations and persist indefinitely in the absence of dissipation.
In the case of $m < \ell$, it is not possible to find a radial function
$Z_{\ell m}(r)$ for which the coupling to both relevant modes, $W_{\ell+1,m}$ and
$W_{\ell-1,m}$ would disappear simultaneously. Therefore, for such modes,
a chain of interactions between toroidal and poloidal modes is set up that
leads to a cascading of flow energy to increasingly higher degrees $\ell$,
i.e., to smaller scales where eventually in a real system dissipation
must take over.

To illustrate this point, we performed numerical simulations with a spectral
transform technique for flow in rotating spherical shells \cite{Christensen15a}.
As initial condition we imposed a single toroidal flow
mode and monitored its evolution.  Momentum advection and viscosity
$\nu$ are retained in the simulations, but are both weak,
with an Ekman number ${\rm E}=\nu/(\Omega d^2)$ = 10$^{-4}$
and a Rossby number ${\rm Ro}=u/(\Omega d)$ of order $10^{-6}$. We set $d$ to 5\% of the radius of the sphere.
Picking the mode $Z_{8,8}$
we find a drift frequency in agreement with the predicted frequency. After the radial dependence has adjusted to the eigenfunction there is negligible energy in modes other than $Z_{8,8}$.  The energy in $Z_{8,8}$ decreases very slowly due to the residual viscosity.  When we start from the mode $Z_{8,7}$ the
evolution is very different. The drift frequency of the basic mode agrees again
with the prediction, but the mode is much more strongly damped.
After eight rotations, the energy in
$Z_{8,7}$ has dropped to 43\% of its original value, whereas in the case of
the $Z_{8,8}$-mode (when started with the correct radial eigenfunction) 99.7\%
of the energy is retained at this time (these numbers depend on the assumed
viscosity). Supplementary Figure~6 compares the radial vorticity distribution
 at time zero with that after eight rotations for the $(\ell, m)=(8,7)$ case.
A range of modes $(\ell,7)$ has been excited and the modes with $\ell \neq 8$ carry collectively 45\% of the
total energy at that point in time. The stronger dissipation at high $\ell$ drains
energy out of the system much more efficiently than it is possible for the basic
mode alone.

These results for a simple physical system provide a plausible explanation
for why we observe only Rossby modes with $m=\ell$ in the Sun, but it
remains to be tested whether the particular stability of such modes also
persists under more realistic conditions such as density stratification or
a free surface.\\

\clearpage

\noindent
{\LARGE \bf Supplementary Information for:}
\vspace{0.5cm}

\noindent
{\Large Global-scale equatorial Rossby waves as an essential component of solar internal dynamics}\\

\noindent
{\Large by B. L\"optien, L. Gizon, A.C. Birch, J. Schou, B. Proxauf,\\ T.L. Duvall Jr., R.S. Bogart, U.R. Christensen}

\vspace{2.cm}
\noindent
\begin{itemize}
\item Supplementary Figure 1: Power spectra of large-scale radial vorticity and horizontal divergence from granulation tracking.
\item Supplementary Figure 2: Vorticity power spectra of sectoral modes from ring-diagram helioseismology.
\item Supplementary Figure 3: Vorticity power spectra for $\ell=m=8$ at three depths in the solar interior.
\item Supplementary Figure 4: Spherical-shell simulation of the evolution of the latitudinally-antisymmetric Rossby mode $(\ell, m)=(8,7)$.
\end{itemize}

\clearpage

\begin{figure}[h]
\centering
\includegraphics[width=1.\textwidth]{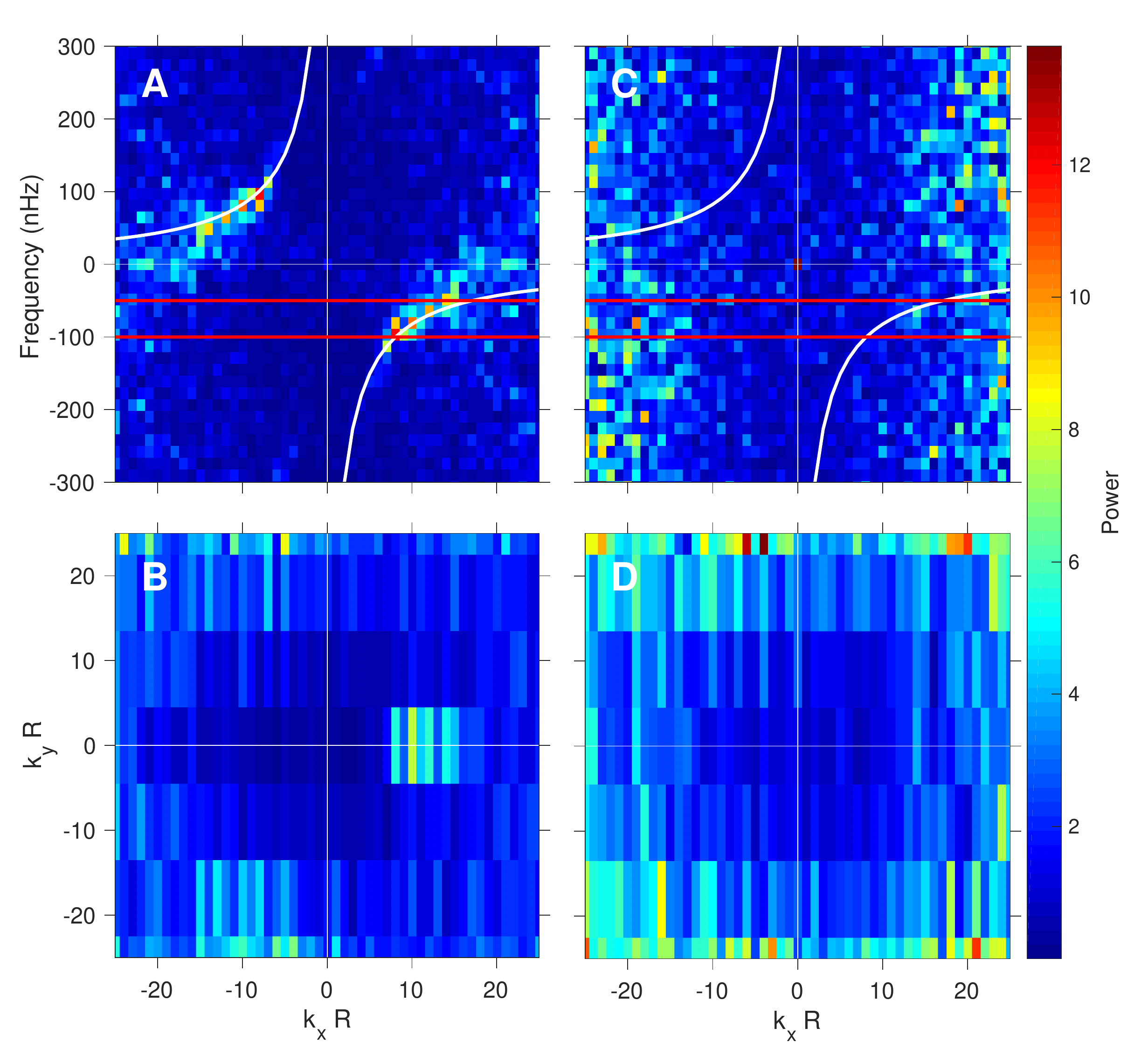}
\caption*{Supplementary Figure 1: {\bf Power spectra of large-scale radial vorticity and  horizontal divergence from granulation tracking.} Power spectra are derived from a strip in latitude (between $\pm 20^\circ$~latitude). Panel~A shows the power of radial vorticity as a function of frequency and local wavenumber in the prograde direction ($k_x$). Panel~B shows the vorticity power averaged over frequencies between $-96.7$~nHz and $-53.7$~nHz (between the red lines), as a function of $k_x$ and $k_y$ , where $k_y$ is the local wavenumber pointing north. Since the data cover only $40^\circ$ in latitude, the resolution of $k_y R$ is only nine. The white curves show the classical Rossby wave dispersion relation. Panels~C and~D show the same configurations for the horizontal divergence.}
\end{figure}

\begin{figure}
\centering
\includegraphics[width=\textwidth]{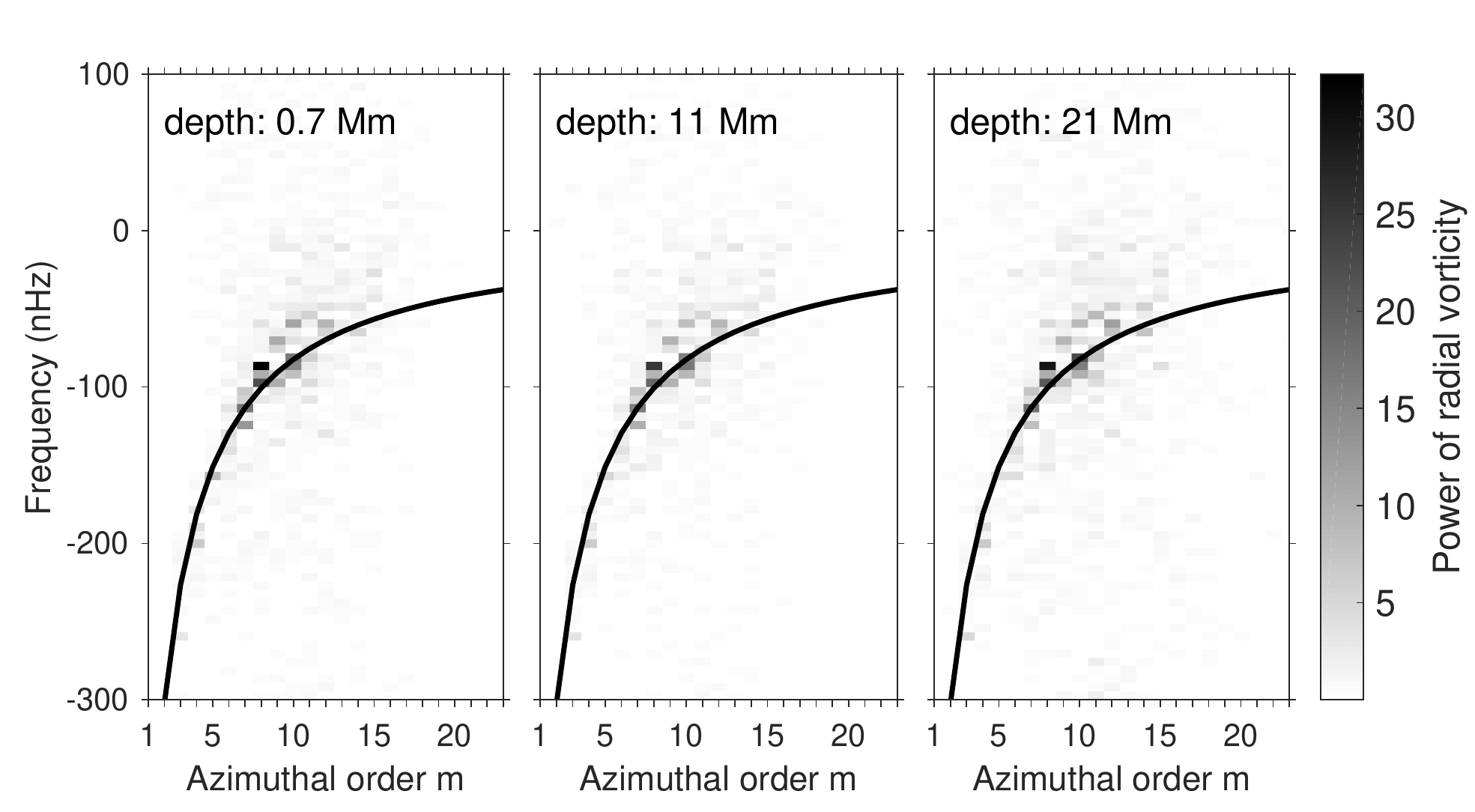}
\includegraphics[width=\textwidth]{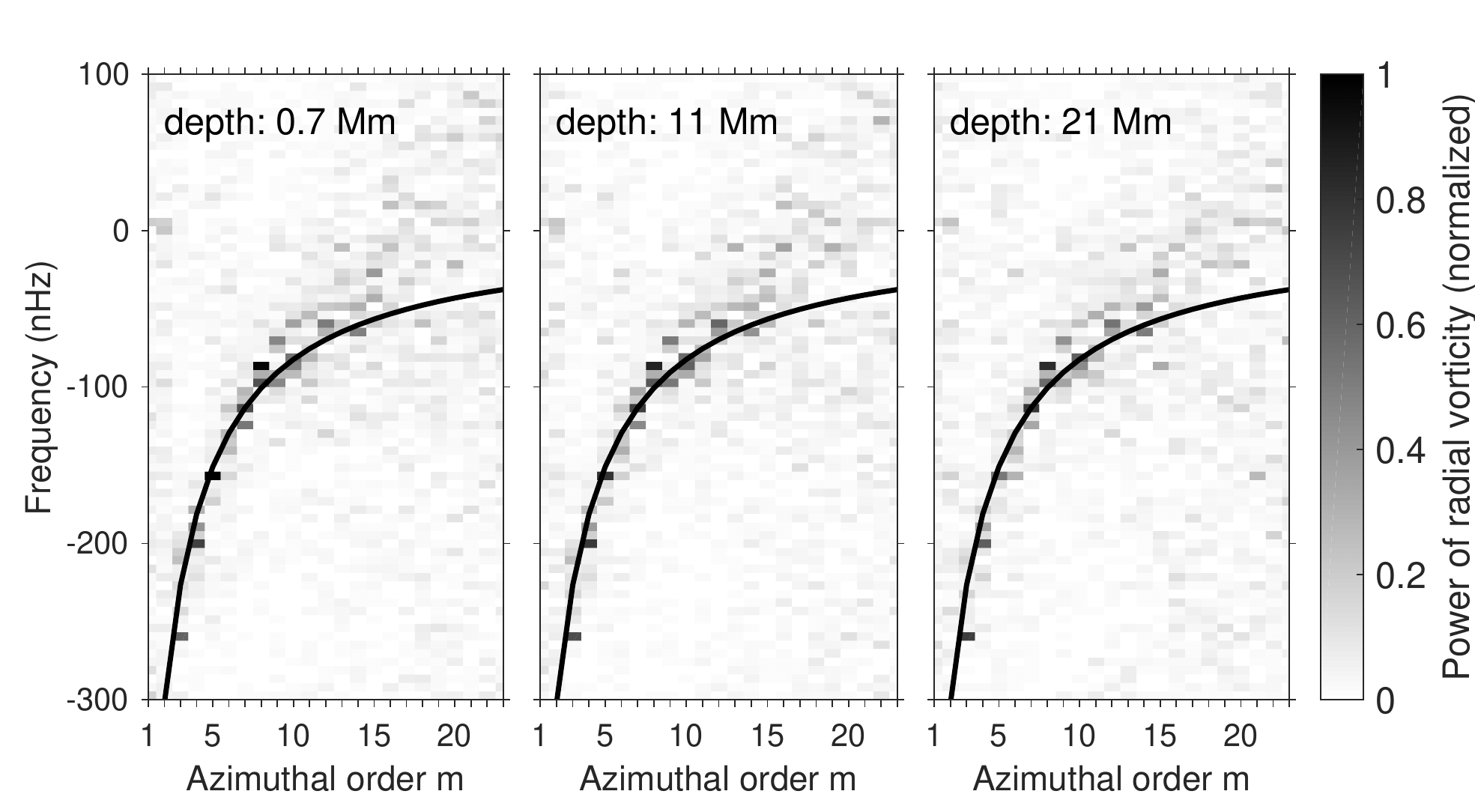}
\caption*{Supplementary Figure 2: {\bf Vorticity power spectra of sectoral modes from ring-diagram helioseismology.} Top panels: Displayed are three power spectra at depths $0.7$~Mm, $11$~Mm, and $21$~Mm. The frequency resolution is 5.4~nHz. The normalization is such that the ring-diagram power in the $m=8$ mode at depth $0.7$~Mm is the same as for the surface data (Fig.~3a). Rossby waves are visible throughout the entire depth range (down to $21$~Mm). Bottom panels: same as top panels except that the power at each $m$ is divided by the power averaged over the frequency range [$-300$, $100$] nHz.}
\end{figure}

\begin{figure}[h]
\centering
\includegraphics[width=0.7\textwidth]{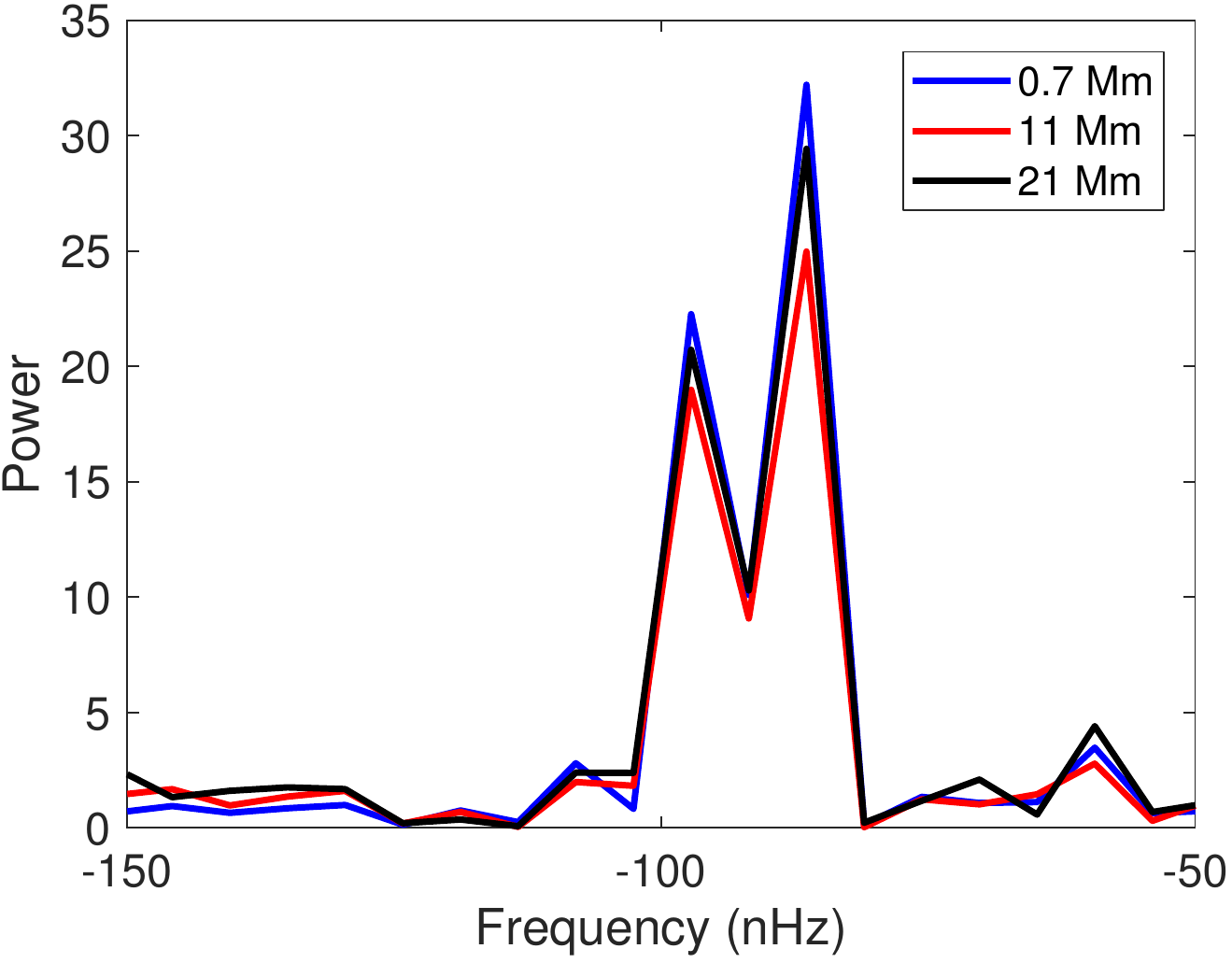}
\caption*{Supplementary Figure 3: {\bf Vorticity power spectra of sectoral modes $\boldsymbol{\ell=m=8}$ at three depths in the solar interior.} 
The Rossby wave amplitudes are similar at depths $0.7$~Mm, $11$~Mm, and $21$~Mm. Furthermore, all three power spectra are similar in shape to the granulation-tracking power spectrum shown in {Fig.~3c}.
}
\end{figure}

\begin{figure}
\begin{center}
\includegraphics[width=10cm]{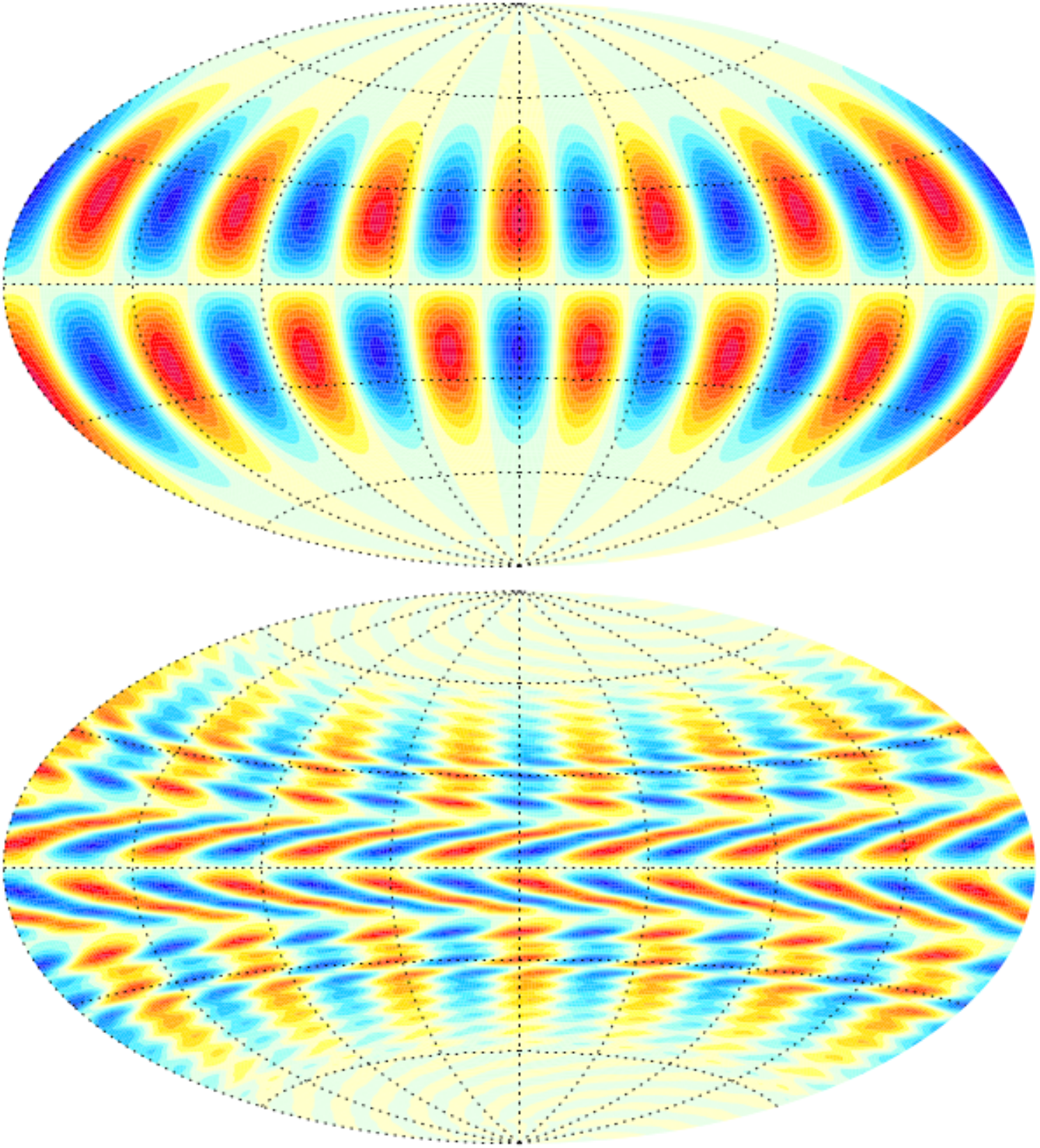}
\end{center}
\caption*{Supplementary Figure 4: {\bf Spherical-shell simulation of the evolution of the latitudinally-antisymmetric Rossby mode $\boldsymbol{(\ell, m) = (8, 7)}$.}
The radial vorticity field is shown at time zero (top panel) and after eight rotations (bottom panel).}
\label{fig:vort87}
\end{figure}

\end{document}